# Continuous Wavelet Transformation and VGG16 Deep Neural Network for Stress Classification in PPG Signals

Yasin Hasanpoor
Advanced Service Robots (ASR) Lab., Department of
Mechatronics Engineering,
Faculty of New Sciences and Technologies,
University of Tehran,
Tehran, Tehran, Iran
yasin.hasanpoor@ut.ac.ir

Bahram Tarvirdizadeh
Advanced Service Robots (ASR) Lab., Department of
Mechatronics Engineering,
Faculty of New Sciences and Technologies,
University of Tehran,
Tehran, Tehran, Iran
bahram@ut.ac.ir

Khalil Alipour
Advanced Service Robots (ASR) Lab., Department of
Mechatronics Engineering,
Faculty of New Sciences and Technologies,
University of Tehran,
Tehran, Tehran, Iran
k.alipour@ut.ac.ir

Mohammad Ghamari
Department of Electrical and Computer Engineering,
Kettering University,
Flint, Michigan, USA,
mghamari@kettering.edu

*Abstract* — **Our research introduces a groundbreaking approach to stress classification through Photoplethysmogram (PPG) signals. By combining Continuous Wavelet Transformation (CWT) with the proven VGG16 classifier, our method enhances stress assessment accuracy and reliability. Previous studies highlighted the importance of physiological signal analysis, yet precise stress classification remains a challenge. Our approach addresses this by incorporating robust data preprocessing with a Kalman filter and a sophisticated neural network architecture. Experimental results showcase exceptional performance, achieving a maximum training accuracy of 98% and maintaining an impressive average training accuracy of 96% across diverse stress scenarios. These results demonstrate the practicality and promise of our method in advancing stress monitoring systems and stress alarm sensors, contributing significantly to stress classification.**

*Keywords*-- **Continues Wavelet Transform (CWT), Photoplethysmogram (PPG), Kalman Filter, Stress Assessment, Visual Geometry Group 16 (VGG16)**

## I. INTRODUCTION

In the vast domain of instrumentation, control, and automation systems, the management of stress has emerged as an increasingly critical concern, transcending its impact on human well-being and extending to the seamless operation of a multitude of systems and processes [1][2]. With the burgeoning demand for real-time stress monitoring and effective stress management strategies, wearable technology has risen as a transformative solution[3][4]. These state-of-the-art wearable devices, equipped with a sophisticated array of sensors, offer the means to monitor and classify mectal stress in real-time, facilitating proactive health management and optimizing the performance of control and automation systems [5][6].

Stress assessment can draw upon various physiological signals, and one that has garnered significant attention for its accuracy and practicality in real-world stress alarm systems is the Photoplethysmogram (PPG) [7][8]. Examining PPG signals enables the detection of blood volume changes, providing valuable information on the cardiovascular system's response to various stress factors [9][10]. Its extensive adoption in the realms of health monitoring and stress alarm systems positions PPG as a dominant signal of interest [11][10]. Its capacity to provide comprehensive insights into the physiological repercussions of stress not only advances our understanding but also lays the foundation for the development of efficacious control and automation strategies [5][3][12]. In this paper, we delve into the application of PPG signals in stress analysis, shedding light on their efficacy in crafting robust stress alarm systems. Our research contributes to the enhancement of instrumentation, control, and automation systems, offering innovative approaches to address the diverse challenges posed by stressors.

In the realm of stress analysis within instrumentation, control, and automation systems, researchers have harnessed physiological signals, such as the PPG, to unveil stress dynamics. While conventional methods like the Fourier Transform (FFT) offer insights into signal frequency analysis, the Continuous Wavelet Transform (CWT) has emerged as a powerful alternative due to its ability to simultaneously explore time and frequency domains, providing a more comprehensive understanding of physiological signal variations under stress conditions [4][7][13]. Additionally, deep learning methodologies, particularly CNNs, have gained prominence for stress assessment [14]. Their proficiency in pattern recognition and feature extraction from complex physiological data contributes significantly to our evolving understanding of stress within our domain, showcasing the convergence of diverse methodologies.

The primary aim of this research is to perform an extensive evaluation of mental stress employing a multi-layered strategy that includes both recognition and classification. The methodology applied in this investigation follows a multi-stage approach to thoroughly evaluate mental stress utilizing physiological signals. To begin, the PPG signals undergo a denoising process utilizing the Kalman filter, enhancing the quality of the data, and making important trends and patterns more prominent [15]. Subsequently, the denoised signals are segmented into windows spanning 10 seconds each with a stride of 1 second, thus facilitating localized analysis.

## II. DATASET DESCRIPTION AND PREPROCESSING

### A. Dataset

We utilized the WESAD dataset as the foundation for our stress assessment research, encompassing data from 15 individuals and providing valuable insights into a range of scenarios [16]. This dataset predominantly focuses on capturing physiological changes through PPG signals. It offers a comprehensive perspective on stress dynamics, spanning various contexts, including calm states, stress-inducing tasks, and meditative intervals. To ensure data authenticity, the dataset relies on wearable sensor technology, faithfully preserving the inherent signal variations that reflect genuine stress reactions[16].

The dataset encompasses a diverse group of participants, including 12 males and several females, with an average age of 27, ensuring demographic variability and enriching our dataset. Data collection devices included the Empatica E4, with a particular focus on PPG signals recorded at a sampling rate of 64Hz. Stress induction employed a widely recognized stress-inducing paradigm known for its effectiveness in triggering stress reactions. Significantly, the sequences involved stress and amusement interspersed with guided meditation exercises aimed at inducing relaxation through controlled breathing. The sequence of signal recordings, depicted in Figure 1, commenced with amusement, followed by meditation, stress, rest, and another meditation phase [16].

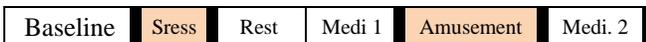

Figure 1. Signal Recordings in Diverse Contexts and Self-Reported Instances within the WESAD Datase

### B. Data Augmentation and Updation

In this study, the data augmentation technique applied a windowing approach to the original PPG signals. Segmentation into 10-second windows facilitated the extraction of CWT representations for each window. By using a stride of 1 second, this approach effectively updated the windows, subsequently generating new CWT representations. As a result, the augmentation process expanded the dataset by re-representing the original signals in a broader scope of windows, each associated with its unique CWT features. This method increased the number of data points by creating multiple CWT representations from the original PPG signals, offering a more comprehensive and varied dataset for subsequent analysis and training of the CNN.

### C. Data Preprocessing

Data preprocessing for our study involves a crucial step to prepare the WESAD dataset for analysis. This dataset provides a wealth of information from 15 participants, including acceleration signals. In order to handle the temporal variability in PPG signals and mitigate data noise, we utilize the Kalman filter. The recursive algorithm is tailored to estimate the state of a linear dynamic system using measurements impacted by noise. The Kalman filter enhances the data quality, ensuring that important trends and patterns in the PPG signals become more prominent [17].

The Kalman filter enhances the data quality by effectively reducing noise in the PPG signals [18]. The filter operates through a two-step process: state prediction and measurement update.

1. State Prediction (a priori):
Predicted State Estimate:

$$X'_k = A \times X_{k-1} + B \times u_k \qquad (1)$$

Predicted Error Covariance:

$$P'_k = A \times P_{k-1} + A^T + Q \qquad (2)$$

2. Measurement Update (a posteriori):
Kalman Gain:

$$K_k = P'_K \times H^T / H \times P'_K \times H^T + R \qquad (3)$$

Corrected State Estimate:

$$X''_k = X'_k + K_k \times (Z_k - H \times X'_k) \qquad (4)$$

Updated Error Covariance:

$$P_k = (1 - K_k \times H) \times P'_k \qquad (5)$$

Which parameters are explained below:

Predicted State Estimate ($X'_k$): This is the estimated state of the system at time k. predicted based on the previous state estimate $X_{k-1}$, a state transition matrix (A), and any control input $B \times u_k$ The predicted state estimate represents the expected value of the state at time k, given the information up to time k-1.

Predicted Error Covariance ($P'_k$) : This is an estimate of the error in the predicted state estimate. It takes into account the previous error covariance ($P_{k-1}$), the state transition matrix (A), and a process noise covariance (Q). The predicted error covariance represents the uncertainty in the predicted state estimate.

Kalman Gain ($K_k$): The Kalman gain is a weight that determines how much the new measurement ($Z_k$) should be used to correct the predicted state estimate. It is computed based on the predicted error covariance ($P'_K$), the measurement matrix (H), and the measurement noise covariance (R).

Corrected State Estimate ($X''_k$): This is the refined estimate of the state at time k, obtained by combining the predicted state estimate ($X'_k$) with the Kalman gain-adjusted measurement residual ($Z_k - H \times X'_k$).

Updated Error Covariance ($P_k$): The updated error covariance represents the reduced uncertainty in the state estimate after incorporating the new measurement. It is computed using the predicted error covariance ($P'_k$), the Kalman gain ($K_k$), and the measurement matrix (H).

The Kalman filter effectively mitigates noise, leading to an improved quality of the PPG signals, which subsequently facilitates more accurate analysis. This denoising procedure highlights significant trends and patterns within the data, making them more prominent. The impact of the Kalman filter on a segment of the PPG signal is visually demonstrated in Figure 2.

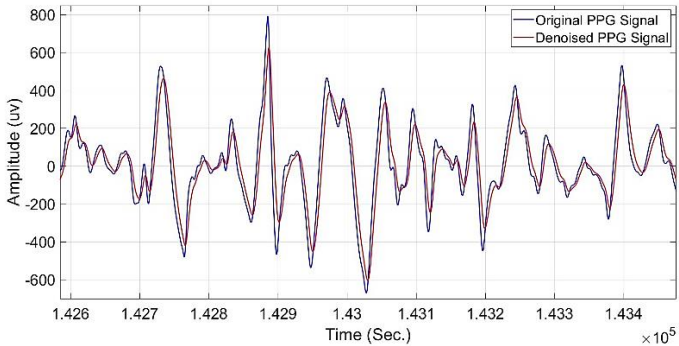

Figure 2. Comparison of PPG Signal Before and After Kalman Filtering

### III. METHODOLOGY

#### A. Utilizing CWT for Image Generation from PPG Signals

In our quest to comprehensively analyze stress through physiological signals, we implement the CWT as a pivotal preprocessing step for PPG signals. CWT serves as a transformative mathematical technique that facilitates the conversion of time-domain PPG data into a format conducive to deep learning analysis [13][19]. The fundamental premise underlying CWT is succinctly summarized by the following mathematical expression of Eqn. 6:

$$CWT(a,b) = \int_{-\infty}^{\infty} x(t) \frac{1}{\sqrt{a}} \varphi\left(\frac{t-b}{a}\right) dt \qquad (6)$$

Within this equation, a and b represent unique coefficients facilitating adjustment and magnification. The variable X(t) denotes our signal at various time instances t, while φ represents a distinctive wavelet function employed to assess the correspondence between our signal and the function.

#### B. Practical Implementation:

Our practical implementation encompasses a series of steps, featuring the application of the VGG16 deep learning classifier:

Data Preparation: We commence by meticulously organizing our PPG signal data within the relevant directories, distinguishing stress and non-stress images. This judicious data organization lays the groundwork for subsequent transformation.

Image Data Generation: We harness TensorFlow and Keras to establish image data generators enhanced with data augmentation techniques. This strategic step empowers us to generate batches of PPG signal images conducive to deep learning analysis.

Model Training: Our chosen deep learning model, VGG16, undergoes rigorous training on the generated PPG signal images. This training regimen extends over a specific number of epochs (in our case, 50 epochs), enabling the model to effectively classify the transformed images. The categorization assigns them to stress and non-stress classes.

Monitoring Progress: The training process's dynamics are visually represented, offering valuable insights into the model's learning evolution. These visualizations encompass the trajectory of accuracy and loss across epochs.

Model Evaluation: A pivotal component of our methodology, model evaluation, is conducted to gauge the approach's efficacy. We critically assess the model's performance through a confusion matrix and a comprehensive classification report, encompassing key metrics tailored to our stress classification task.

Model Architecture Visualization: To facilitate a deeper understanding, we visually depict the architecture of our VGG16-based model. This graphical representation provides a transparent view of the model's structural elements and connections.

## C. VGG16-Based Classifier Structure

The core of our stress classification methodology is the utilization of a VGG16-based classifier. This section offers insights into the architectural intricacies of this classifier and its role in our stress analysis pipeline. The VGG16-based classifier consists of multiple convolutional and fully connected layers that enable hierarchical feature extraction. It comprises 16 weight layers, hence the name "VGG16." The architecture can be succinctly summarized as follows:

Convolutional Blocks: VGG16 comprises multiple convolutional blocks, each containing convolutional layers with small 3x3 filters. These blocks are interspersed with max-pooling layers for spatial down-sampling.

Dense Layers: Positioned at the latter part of the network, the dense layers intricately analyze the high-level features extracted by the preceding convolutional layers. This strategic arrangement serves as a critical component for conducting detailed feature analysis.

Fully Connected Layers: Situated towards the network's conclusion, these fully connected layers meticulously examine the high-level features extracted by the preceding convolutional layers. This configuration plays a crucial role in conducting in-depth feature analysis.

In summary, the VGG16-based classifier is a pivotal component of our stress analysis framework, leveraging its deep architecture for effective feature extraction and classification. Its intricate structure, as depicted in Figure 3, underscores the model's capacity to discern subtle patterns in PPG signal images, thereby enabling accurate stress classification. This advanced model excels in discerning intricate patterns within PPG signal images, ensuring accurate stress identification within the given stress and non-stress categories.

## IV. RESULTS AND CONCLUSION

The training and validation process played a pivotal role in honing the performance of our stress classification framework. Figure 4 visually represents the progression of this process, displaying the model's accuracy and loss metrics across 50 epochs.

The VGG16 model undergoes training utilizing a categorical cross-entropy loss function, a prevalent choice for tasks involving classification. In the context of our two-class problem (stress vs. non-stress), This loss function quantifies the disparity between the predicted class probabilities and the true class labels, penalizing the model for making incorrect predictions. Alongside the loss function, the key hyperparameter set includes an initial learning rate of 1.0e-05, which influences the rate at which the model adjusts its internal parameters to optimize the loss function. The accuracy metrics, both for training and validation sets, are reported to assess the model's performance on both seen and unseen data. Throughout the training process, the loss and accuracy metrics fluctuate across epochs, reflecting the model's learning process and performance on the given stress classification task.

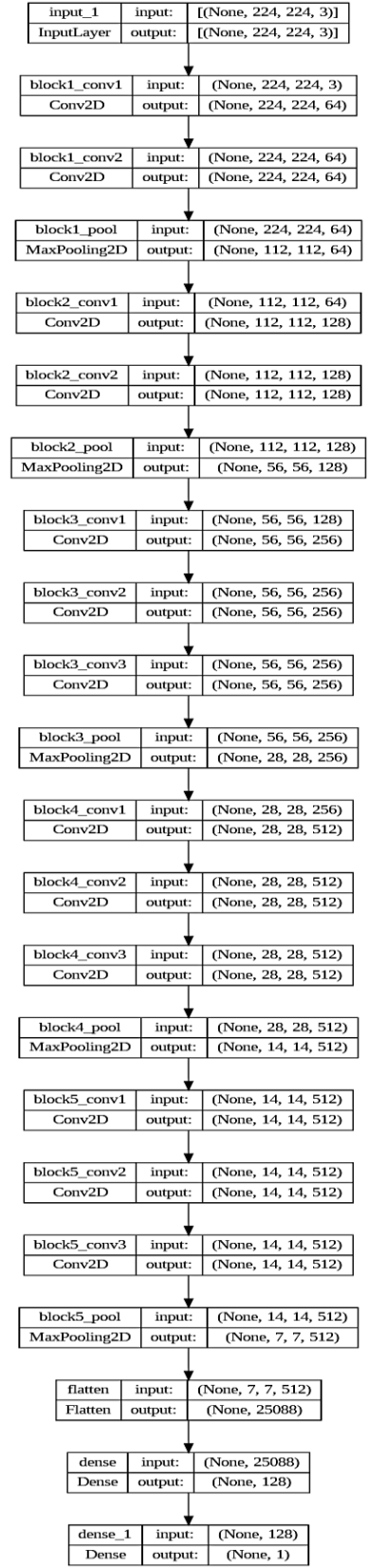

Figure 3. Visual Structure of the VGG16-Based Classifier

The model exhibited impressive learning capabilities from the outset. In the initial epochs, there was a substantial increase in accuracy, with the training dataset showing an accuracy of approximately 97% while the validation dataset achieved around 95% accuracy. This early success suggests that our system is effectively learning and extracting key features from the continuous wavelet-transformed PPG signal images.

However, as the training progressed, we observed some fluctuations in performance. While the training accuracy continued to improve, gradually reaching above 98%, the validation accuracy experienced minor variations, occasionally dipping below 94%. These variations could be explained by the model's flexibility and its ability to manage diverse stress patterns.

The training and validation losses followed a similar trend. The training loss consistently decreased, indicating that the model was minimizing errors as it learned from the data. On the other hand, the validation loss showed occasional upward spikes, signifying challenges in generalizing to unseen stress patterns.

A significant achievement was the model's ability to maintain relatively high accuracy, even in the later stages of training. The convergence of training and validation accuracy and loss suggests that the model was robust and resistant to overfitting. This robustness is particularly important as it ensures the generalization of our stress classifier to new and unseen data, a crucial attribute for a stress management system in real-world scenarios.

The slight variations observed in validation accuracy and loss are common in machine learning models and may result from the inherent variability in physiological stress responses among individuals. These fluctuations emphasize the importance of a diverse dataset that encompasses a wide range of stress scenarios and individuals.

In conclusion, the training and validation process of our VGG16-based classifier demonstrated a highly capable system for stress classification. While some variations were observed, the model maintained robust accuracy, indicating its potential for real-world applications in stress management and wearable sensors. Further exploration is needed to investigate the performance on larger and more diverse datasets, potentially uncovering new insights and strategies to improve its overall stability and accuracy.

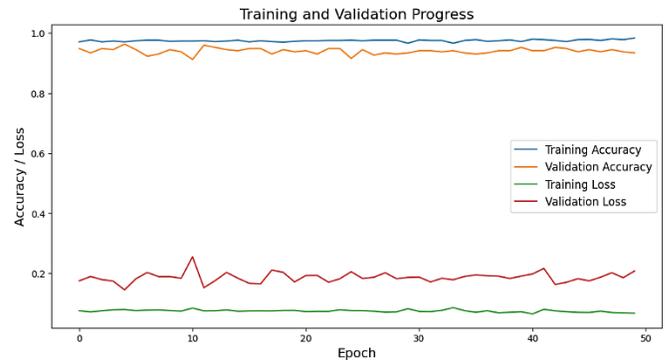

Figure 4. Training and Validation Progress of the VGG16-based Stress Classifier